# Ferrimagnetic Heusler tunnel junctions with fast spin-transfer torque switching enabled by low magnetization

Chirag Garg[1+*], Panagiotis Ch. Filippou[1*], Ikhtiar[3*], Yari Ferrante[1], See-Hun Yang[1], Brian Hughes[1], Charles T. Rettner[1], Timothy Phung[1], Sergey Faleev[1], Teya Topuria[1], Mahesh G. Samant[1+], Jaewoo Jeong[3+,], and Stuart S. P. Parkin[2+]

[1]IBM Research - Almaden, San Jose, California 95120, USA.

[2]Max Plank Institute for Microstructure Physics, Weinberg 2, 06120 Halle (Saale), Germany.

[3]Samsung Semiconductor, Inc., San Jose, California 95134, USA

* These authors contributed equally to this work.

+ Emails of corresponding authors: chirag.garg1@ibm.com, j.jeong1@samsung.com, stuart.parkin@mpi-halle.mpg.de, mgsamant@us.ibm.com

**Magnetic random access memory that uses magnetic tunnel junction memory cells is a high performance, non-volatile memory technology that goes beyond traditional charge-based memories. Today its speed is limited by the high magnetization of the memory storage layer. Here we show that fast and highly reliable switching is possible using a very low magnetization ferrimagnetic Heusler alloy, $Mn_3Ge$. Moreover, the tunneling magnetoresistance is the highest yet achieved for a ferrimagnetic material at ambient temperature. Furthermore, the devices were prepared on technologically relevant amorphous substrates using a novel combination of a nitride seed layer and a chemical templating layer. These results show a clear path to the lowering of switching currents using ferrimagnetic Heusler materials and, therefore, to the scaling of high performance magnetic random access memories beyond those nodes possible with ferromagnetic devices.**

Spintronic magnetoresistive devices have powered the reading elements in hard disk drives for more than two decades, thereby enabling a ~10,000 fold increase in storage capacity. The original spin valve devices that were introduced in 1997[1] were based on spin-dependent interfacial resistive scattering[2,3]. These were superseded by devices that are based on spin-dependent tunneling[4-7]. These latter devices were also proposed for solid-state non-volatile magnetic memory cells as early as 1995. The first demonstration was made in 1999 when very fast reading and writing (~10 ns) was demonstrated in magnetic tunnel junctions (MTJs) that used amorphous $Al_2O_3$ tunnel barriers, which exhibited up to ~40% tunneling magnetoresistance (TMR)[8-10]. A major breakthrough in 2004 was the demonstration of much higher TMR in MTJs that used crystalline MgO(100) tunnel barriers in conjunction with *bcc* CoFe (or amorphous CoFeB) electrodes[11,12]. Subsequently, MTJs with thinner CoFeB-based electrodes that exhibited perpendicular magnetization were shown to reduce the electrical current that is used to write the memory state by the phenomenon of spin-transfer torque (STT)[13-16]. These materials have since been incorporated into MTJs, forming the basis of recently introduced embedded MRAM products[17] at the 28 nm node. The speed, however, is limited by the use of the relatively high magnetization CoFeB material. This is due to the spin-transfer angular momentum conservation[18-20]. Replacing the CoFeB storage layer with a magnetic material that has a significantly lower magnetization would enable a wider use, in particular for cache memories, that require much higher speeds. At the same time the smaller the current needed, the smaller can be the access transistor, thereby enabling higher densities.

Some longstanding challenges have prevented the adoption of low magnetization materials. While many ferromagnetic materials exhibit low magnetization, most of them have low Curie temperatures (<400 K)[21,22]. A class of ferrimagnets composed of Rare-Earth elements (Gd, Tb) and transition metals (Co, Fe), while easy to grow, are unsuitable not only owing to the low Curie temperature of the Rare-Earth component, but also because of their low thermal structural stability[23,24]. In contrast, certain ferrimagnetic materials such as $Mn_3Ga$[25] and $Mn_3Ge$[26] have both low magnetization and much higher Curie temperatures, thanks to their compensated yet strongly exchange coupled spin structure. The challenges with these compounds are the growth of ultra-thin layers with bulk-like properties on technologically relevant substrates as well as achieving high TMR at a low enough resistance-area (RA) product which is suitable to facilitate STT-switching[27]. In such ferrimagnets, the spin polarization of the tunneling current may be compromised due to the compensated nature of the two antiferromagnetically coupled Mn sub-lattices, especially at the interface[28].

In our current work, we address these issues to demonstrate an STT-switchable MTJ formed from $Mn_3Ge$, a tetragonal ferrimagnetic Heusler alloy with low magnetization[29-33], with high TMR and a low RA = 11.4 $\Omega\mu m^2$. We find these MTJs have much lower switching currents at high write speeds than for MTJs formed from conventional ferromagnetic electrodes for otherwise the same thermal stability. Moreover, these $Mn_3Ge$-based MTJs are formed on amorphous $SiO_x$ and are thermally stable at temperatures above 400 °C making them compatible with typical CMOS back-end of line processing. A particular advantage of Heusler ferrimagnets is that the two magnetic sub-lattices are formed from transition metals and thus the net moment is weakly dependent on temperature for the operational temperature window of the MTJ, in contrast to, for example, rare-earth – 3d transition metal ferrimagnetic alloys whose net magnetization varies considerably with temperature[34]. This makes tetragonal ferrimagnetic Heusler alloys ideal candidates for high-speed magnetic memory applications.

A recently developed chemical templating layer (CTL) technique[35,36] was used to grow the $Mn_3Ge$-based MTJ stacks but here we refine the technique to allow for growth on Si/SiO$_x$ substrates, which is required for CMOS compatibility (see below). The film stacks were patterned into circularly shaped devices, 30-40 nm in diameter, using electron-beam lithography and Ar ion-beam milling. A typical ~36 nm diameter device is shown in the bright-field transmission electron microscopy (BFTEM) cross-sectional image presented in Fig. 1a. The $Mn_3Ge$ forms the lower electrode, i.e. the free layer (FL). The top electrode, the reference layer (RL), is formed from CoFeB that is ferromagnetically exchange-coupled, via a thin Ta spacer layer, to a synthetic antiferromagnetic (SAF) structure composed of two distinct [Co/Pt] multilayers separated by a thin Ir layer[37,38]. The upper of these [Co/Pt] multilayers is designed to have higher moment than the lower CoFeB/Ta/[Co/Pt] multilayers of the film structure. The two memory states of the MTJ correspond to the moment of the $Mn_3Ge$-FL being parallel (P) or antiparallel (AP) to the moment of the lower layer of the RL, with corresponding resistance values, $R_P$ and $R_{AP}$, respectively.

Figures 1b-c show the switching between the P and AP states of the MTJ, driven by field (b) and current (c). The MTJ is switched reversibly by using an out-of-plane (OOP) magnetic field (Fig. 1b) resulting in two well defined, non-volatile states in the absence of an external magnetic field. By initializing the net moment of the SAF along $+z$, we infer from the magnetic hysteresis loop that the higher resistance belongs to the AP configuration. Therefore, the resulting TMR ($\frac{R_{AP}-R_P}{R_P}$) is positive, exhibiting a value of +45%. Interestingly, the sign of the TMR was found to be negative for thicker $Mn_3Ge$ films in an earlier study[28]. In Fig. 1c, the spin transfer-torque driven switching of this device is demonstrated by applying voltage pulses of increasing amplitude. Reversible current induced switching back and forth between the P and AP configurations is clearly observed. From the resistance versus voltage (R-V) hysteresis loop, it can be confirmed that the $Mn_3Ge$-FL is switching.

CTLs are binary compounds with the CsCl structure which have been shown to promote chemical ordering in ultra-thin Heusler alloy films deposited on them[35] even at single unit cell thicknesses and at ambient temperature. They also enable the desired perpendicular magnetic anisotropy (PMA) in the Heusler layer[35]. This previous work used single crystalline MgO (001) substrates. In order to promote the growth of the CTL on amorphous $SiO_2$ layers we carried out a wide-ranging materials exploratory search. We found that binary nitrides with a NaCl structure readily grow with a (001) texture on such amorphous surfaces. The nitrides TiN, VN, CrN, TaN, MnN and ScN are all effective for this purpose (see Supplementary Table 1) and allow for a wide range of lattice constants for epitaxial matching with the CTL. Two of these nitrides, MnN (see supplementary Figure 1) and ScN, were chosen for further study. It was determined that metallic MnN is not as thermally stable as semiconducting ScN at ~400 ºC, the annealing temperature required for back-end-of-line (BEOL) integration.

Cross-sectional high-angle annular dark-field scanning transmission electron microscopy (HAADF-STEM) images that are presented in Figs. 2a and 2b show the epitaxial growth of a $Mn_3Ge$ layer for an MTJ stack grown on $Si/SiO_2$. The CTL was grown over a ScN seed layer, here only ~1 Å thick. The detailed structure of the MTJ is as follows $Si/SiO_2$/ 50Ta/ 5CoFeB/ 1ScN/ 400Cr/ 50IrAl/ 150CoAl/ 19$Mn_3Ge$/ 14MgO / 14.5CoFeB/ 50Ta/ 100Ru (thickness values in angstroms). A 400Å Cr layer is included to allow for Current In-Plane Tunneling (CIPT) measurements. The CTL is formed from a bilayer of IrAl and CoAl, both of which exhibit a CsCl structure with a (001) texture, thereby templating the growth of a chemically ordered $Mn_3Ge$ (001) layer on top (see Fig. 2b). In particular note that the $Mn_3Ge$ exhibits alternating ferromagnetically aligned layers of Mn-Mn and Mn-Ge whose magnetizations are coupled antiferromagnetically[39]. This structure gives rise to a significant volume perpendicular magnetic anisotropy (PMA) along the tetragonal (001) axis[30]. A CTL from a single layer of CoAl also is effective but we found that higher TMR was possible by using the bilayer CTL. Similarly, ScN can be replaced by other nitrides (Supplementary Table 1). To further demonstrate the role of ScN we present x-ray diffractograms (XRDs) of the above stack and the related stack without any nitride layer ($Si/SiO_2$/ 50Ta/5CoFeB /400Cr /150CoAl /19$Mn_3Ge$ /14MgO /20Ta) in Fig. 2c. In the former case well defined Cr (002) and CoAl (001) peaks are observed, whereas in the latter case no such peaks are found. This comparison shows that the CoAl CTL has the desired (001) texture only when grown on nitride or nitride/Cr sublayers.

MTJs grown in this way show $Mn_3Ge$ free layers with the desired magnetic properties, as seen in the resistance (measured by CIPT) versus out-of-plane field (R-H) hysteresis loop in Fig. 2d. CIPT-TMR measured in this MTJ stack is 69% but when patterned into devices gives rise to higher TMR values (see Supplementary Figure 2). Note that the TMR is positive, i.e. $R_{AP}>R_P$, which is opposite to previously reported results for MTJs with much thicker (300 Å) $Mn_3Ge$ layers grown without a CTL (TMR~-35%)[28]. We also find negative TMR values for thick (50Å) $Mn_3Ge$ layers grown using CTL layers with TMR values as high as ~-109% (see

Supplementary Figure 3) but these layers were not current switchable due to their very high magnetic anisotropy. Nevertheless, this is the highest TMR value yet reported for any ferrimagnetic material.

The sign difference in TMR for strained and relaxed Mn3Ge layers reflects the larger in-plane lattice constant and reduced tetragonality of the Mn3Ge layer for the MTJ grown with the CTL. Indeed, the TEM images (Figs. 2a and 2b) show that the thin Mn3Ge layer assumes the in-plane lattice parameter (~4.03Å) of the CoAl CTL. A positive TMR was found from Density Functional Theory (DFT) calculations (see Supplementary Figure 4) for MTJs that include a Mn3Ge layer with the same in-plane lattice constant as that found here for the CoAl CTL. For Mn3Ge layers with the bulk lattice constant previous DFT calculations show a negative TMR[28]. Note that the TMR sign is determined by which of the two Mn-Mn or Mn-Ge ferromagnetic layers have the larger magnetic moment which in turn depends on the in-plane tensile strain according to DFT calculations. Furthermore, the spin polarization of the tunneling current depends sensitively on this strain so that, surprisingly, the spin polarization from each of the Mn layers can have the same sign and, therefore, higher TMR. Indeed, our DFT calculations show the possibility of very high TMR values of up to +400 % (as discussed in Supplementary Figure 5). The largest TMR that we find for current switchable Mn3Ge layers is +87 % (Supplementary Figure 2).

The magnetic properties of Mn3Ge thin films were studied by growing stacks identical to MTJ stacks but without the RL. The magnetic OOP hysteresis loops for a series of Mn3Ge thin films with varying thickness $t_{Mn3Ge}$ are shown in Fig. 3a. All the films exhibit PMA with high remanence. We find that the magnetization ($M_s$) and coercivity ($H_c$) are very sensitive to small changes in $t_{Mn3Ge}$ (Fig. 3b). $H_c$ more than doubles for every 2Å increase in thickness reaching a very high value of ~42 kOe for $t_{Mn3Ge}$ = 21 Å. Being able to attain such high values of $H_c$ can be beneficial for making devices impervious to external fields. Thicker films exceed 70 kOe, the maximum field in our measurement apparatus. For $t_{Mn3Ge}$ = 11 Å, $M_s$ ~165 emu/cc, the bulk value[33], but decreases as $t_{Mn3Ge}$ is further increased. This observation likely arises from a subtle shift in the delicate balance of magnetic moments in the MnGe and Mn layers. The $H_k$ values were extracted from in-plane magnetic hysteresis loops (see Fig. 3c). For $t_{Mn3Ge}$>17 Å, $H_k$ was too large (>70 kOe) to be measured (see Supplementary Note 3).

A useful parameter is the thermal stability factor that is given by $\Delta = \frac{E_B}{kT}$ where $E_B$ is the energy barrier to switch the magnetic FL volume ($V_m$) between the P and AP states and $k$ is the Boltzmann's constant. For MRAM applications Δ should exceed ~50 to meet data retention requirements. For single magnetic domain reversal (macrospin approximation), $\Delta = \frac{M_s V_m H_k}{2kT}$ but typically smaller values are found, except in small devices (<30 nm)[40,41]. Experimentally, $\Delta$ can be extracted from STT-driven MTJ switching experiments when the switching voltage $V_{sw}$ is less than $V_{C0}$, the threshold voltage for switching in the absence of thermal fluctuations[42,43]. $V_{sw}$ is extracted from R-V scans (see Fig. 1c for the case of 0.5 ms long pulses). $V_{sw}$ is plotted versus the voltage pulse length ($t_p$) ranging from 0.5 to 100 ms in Fig. 3e for a 30 nm diameter device with $t_{Mn3Ge} = 17$ Å. Within the macrospin approximation[42-44], the slope of $V_{SW}$ vs $\ln\left(\frac{t_p}{\tau_0}\right)$ gives $\frac{1}{\Delta}$. Values of $\Delta$ thus obtained for both P→AP and AP→P switching processes is then averaged and a value of ~60 is estimated for this device. The mean $\Delta$ averaged over a set of 30 to 60 devices with a diameter of 35 nm is plotted versus $t_{Mn3Ge}$ in Fig. 3f. $\Delta$ increases, as would be expected, as $t_{Mn3Ge}$ is increased. Moreover, the values are in good agreement with estimates based on the values of $M_s$ and $H_k$ deduced from Fig. 3b,d. Our results show that Mn$_3$Ge-FLs may enable scaling the diameter of such MTJs to below ~10 nm.

Having established that Mn$_3$Ge-FLs have high thermal stability from millisecond STT-switching characteristics, we now discuss their current-driven switching properties at much shorter time scales (≤10 ns), where thermal fluctuations play little role. Instead, the reversal takes place above the threshold current ($I_{C0} = \frac{V_{C0}}{R_p}$) at which the STT surpasses the intrinsic damping torque[45,46]. In this precessional regime the switching becomes faster with increasing current. At a given temperature, the reversal starts from a thermally distributed initial state. The switching current $I_c$ for a given write-error rate ($WER$) is given by the following equation[19,44,46]:

$$I_C = \frac{4e}{\mu_B g P}\left(\alpha \gamma E_B + \frac{M_s V_m}{4t_p} \log\left(\frac{\pi^2 \Delta}{4WER}\right)\right) \quad (1)$$

or alternatively expressed as

$$I_C = I_{C0} + I_{\text{overdrive}} \quad (2)$$

Here, $\mu_B$, $g$, $P$, $\alpha$, $\gamma$ are the Bohr magneton, g-factor, spin-polarization, Gilbert damping constant and gyromagnetic ratio, respectively. In the long-pulse limit $I_C$ reaches a saturation lower bound value that is given by $I_{C0} = \frac{4e\alpha\gamma E_B}{\mu_B g P}$ and which is governed by $E_B$ and $\alpha$. The efficiency of the switching process is often characterized by the term $\frac{\Delta}{I_{C0}}$, which should be as high as possible. $\frac{\Delta}{I_{C0}}$ can be increased, for example, by suppression of spin-pumping by using an MgO dielectric cap layer on top of the FL[47,48] or making use of two tunnel barriers[49] to increase the spin torque. In the short-pulse limit, the switching current is increased above the saturation lower bound value by the term $I_{\text{overdrive}} = \frac{4e}{\mu_B g P} \frac{M_s V_m}{4 t_p} log\left(\frac{\pi^2 \Delta}{4 WER}\right)$. This term scales with $t_p^{-1}$ and the magnetic moment of the free layer.

It is clear from Eq. (1) that a high $M_s$ is detrimental for current-driven switching at short pulses although, for typical ferromagnets, it contributes towards a high $E_B$. In contrast, ferrimagnets that have a low $M_s$ and high $H_k$, such as Mn$_3$Ge used here, can attain a high $E_B$ whilst reducing $I_{\text{overdrive}}$. We now consider the overdrive term by studying the ratio $\frac{J_C}{J_{C0}}$ where $J_C$ is the switching current density and $J_{C0}$ is the threshold saturation lower bound current density. For a given $t_p$ Eqn. (1) can be rewritten as:

$$\frac{J_C}{J_{C0}} = \left[1 + \frac{\tau_D}{t_p}\left(\frac{1}{2} log\left(\frac{\pi^2 \Delta}{4 WER}\right)\right)\right] \qquad (3)$$

Here $\tau_D$ is the characteristic timescale of switching, that equals $\frac{1}{\alpha \gamma H_k}$. Calculated values of $\frac{J_C}{J_{C0}}$ are plotted in Fig. 4a as a function of $H_k$ using Eq. (3). Note that as $H_k$ is varied, $M_s$ is correspondingly adjusted to maintain the value of $\Delta$ to be 60. The value of $\alpha$ was set at 0.01. For longer pulse lengths (~10 ns) $\frac{J_C}{J_{C0}}$ is small and relatively insensitive to $H_k$. As clearly shown in Fig. 4a, for low $H_k$ and short $t_p$, $J_C$ is calculated to increase to more than an order of magnitude higher than $J_{C0}$ but, on the other hand, ferrimagnetic layers with a low $M_s$ and high $H_k$ will dramatically reduce $\frac{J_C}{J_{C0}}$ even at sub-nanosecond speeds.

As shown in Eqn. 3 there is a finite probability that a device will not switch for a given intensity and length of the current pulse that is given by the parameter WER. The WER was determined by first setting the MTJ to either the AP or P state using an appropriate current pulse and then applying a switching current pulse. The state of the MTJ was then read with a much smaller current sensing pulse. This procedure was then repeated up to $10^7$ times for current pulses with varying magnitudes and durations. Results for the same Mn$_3$Ge-FL device as in Fig. 3e are shown in Fig. 4 b and c. Fig. 4b shows how $\frac{J_C}{J_{C0}}$ varies as the $t_p$ is increased at a fixed $WER = 0.5$ for switching from P to AP and AP to P. Experimental data for both switching processes are fitted with Eqn. 3 to obtain $J_{C0}$ and $\tau_D$. These values are shown in the Figure. Based on the fitted data the STT switching efficiency $\frac{\Delta}{I_{C0}}$ is 1.37. This value is comparable to those reported in the literature for conventional ferromagnetic materials[16,50]. For 1 ns long current pulses we observe the increase in $J_C$ over $J_{C0}$ to be 38 % (at WER=0.5). For MTJs with CoFeB FLs this increase is typically more than 200 %[51,52]. This is simply due to the large difference in $M_s$ for these two materials.

The values of WER are plotted versus $J_C$ for several values of $t_p$ in Fig. 4c. To obtain lower WER values $J_C$ must be increased. Fits to the WER vs $J_C$ plots for each $t_p$ are shown as solid lines (see Methods). For some $J_C$ no errors were detected (hollow points) for the maximum number of switching attempts ($10^7$) so they are shown at $WER = 10^{-7}$. From these data a plot of $\frac{J_C}{J_{C0}}$ vs $t_p$ for $WER = 10^{-6}$ is shown in Fig. 4d. The values of $\frac{J_C}{J_{C0}}$ agree well with those calculated using the value of $\tau_D$ obtained from the data in Fig. 4b. In particular $\frac{J_C}{J_{C0}}$ for 1 ns long current pulse at WER = $10^{-6}$ is ~1.85 which is again considerably smaller than that reported for conventional ferromagnetic free layers[51].

In summary, we find that the switching current needed to set the state of an MTJ memory element at high speeds is significantly lowered by using a ferrimagnetic Heusler free layer that has a very low saturation magnetization compared to the ferromagnetic CoFeB-based free layers used in today's MRAM products. Moreover, these MTJ devices that were prepared on CMOS compatible amorphous substrates show reliable switching up to very low write error rates as well as high TMR, high thermal stability, and high field immunity. The low switching currents that we have demonstrated will allow the use of minimum size drive transistors, thereby, enabling a substantial reduction in the footprint of the memory cell and overcoming a fundamental roadblock to the widespread application of spintronic technologies based on MTJ devices.

**Figure captions:**

**Figure 1**

**Stack details and switching characteristics of a Mn₃Ge-FL MTJ device. a,** BFTEM (bright-field transmission electron microscopy) cross-section image of a Mn₃Ge-FL MTJ device with a diameter of ~ 36 nm. The MTJ film stacks had the following structure: Si(001)/ 250 SiO$_2$/ 50 Ta/ 5 CoFeB/ 1ScN / 400 Cr/ 100 ScN / 10 CoAl/ 17 Mn$_3$Ge (in-situ annealed at 355 °C)/ 13 MgO/ 13.5 CoFeB/ 2.4 Ta/ 6 Co/ 10 Pt/ [2.5 Co/ 5 Pt]$_{x3}$/ 7 Co/ 5.7 Ir/ 6Å Co/ 5.5 Pt/ [5.5 Co/ 5 Pt]$_{x4}$ / 100 Ru. Thickness values are in angstroms. For scale, a white horizontal bar near the bottom corresponds to 10 nm. The bottom and top electrode are connected to signal, ground respectively. **b**, $R$ vs. (out-of-plane) $H$ **c**, and $R$ vs $V$ where the length of applied voltage pulse is 0.5 millisecond. $R$ is measured with a small bias of 50 mV. Note that a static magnetic field equal to the offset field derived from the R-H loop is applied for the R-V measurements to compensate for the fringing dipole field from the top electrode.

**Figure 2.**

**Growth of crystalline chemical templating layer on silicon substrates. a,** HAADF cross-sectional STEM image of a representative Mn$_3$Ge-based MTJ stack grown on silicon substrate. The stack consists of: Si/SiO$_2$/ 50Ta/ 5CoFeB/ 1ScN/ 400Cr/ 50IrAl/ 150CoAl/ 19Mn$_3$Ge/ 14MgO / 14.5CoFeB/ 50Ta/ 100Ru. **b**, high resolution HAADF imaging of the indicated region in a, showing the highly epitaxial growth attained even on silicon substrates. **c**, Out-of-plane $\theta$-$2\theta$ XRD scans of the stack shown in (a) (in red), showcasing the CsCl structure of CoAl CTL as seen from the (001) and (002) peaks as indicated and the exemplary stack without any nitride layer (in black). **d**, CIPT R-H loop showing high TMR from Mn$_3$Ge grown on ScN/CoAl CTL on silicon substrate. The white scale bars in (a-b) correspond to 2 nm.

**Figure 3.**

**Magnetic measurements and estimation of $E_B$ for various $t_{Mn_3Ge}$. a,** out-of-plane $M$-$H$ curves for Mn$_3$Ge-FL film without any RL. **b,** Extracted $M_s$ (black) and $H_c$ (red) from $M$-$H$ curves. Blue dashed line is reference for bulk Mn$_3$Ge $M_s$. **c,** saturation magnetic moment, $m_s$ normalized to a sample area of 0.2 cm$^2$. **d,** $H_k$ extracted from in-plane $M$-$H$ curves (see Supplementary Note 3). **e,** Linear fit to the variation of switching voltage ($V_{SW}$) with $\ln\left(\frac{t_p}{\tau_0}\right)$ as per the macrospin model for a device with diameter ~30 nm and $t_{Mn_3Ge}$ = 17 Å. $\tau_0$ = 1 ns. $V_{SW}$ is calculated as an average from 20 repeated measurements. **f,** experimentally obtained $E_B$ (black) from a set of 35 nm diameter devices and theoretically calculated estimate (red) based on $M_s$ and $H_k$ from 3b,d. Error bars in the figure correspond to one standard deviation. The 35 nm group includes any devices in the size range 32.5 – 37.5 nm.

**Figure 4.**

**Scaling of switching currents at high speeds. a,** $\frac{J_C}{J_{C0}}$ calculated for WER = 0.5, $\Delta$ = 60 and $\alpha$ = 0.01 using Eq. (3). **b,** $\frac{J_C}{J_{C0}}$ vs $t_p$ for Mn$_3$Ge-FL MTJ with $t_{Mn_3Ge}$ = 17 Å (same device from Fig. 3e). Measurement data for P-AP (blue) and AP-P (red) and the average (black) is shown using fits to Eq. (3). **c,** Measured $J_C$ vs WER for different $t_p$ (solid points) with fits to the WER (solid lines) for the same device. The open symbols are measurements where no error was detected for $10^7$ switching events and a WER of $10^{-7}$ was assumed as the upper bound. **d,** $\frac{J_C}{J_{C0}}$ from **c** (black) and based on the Eq. (3) estimate (red) are plotted for different $t_p$.

**Acknowledgements**
We thank Holt Bui, Eugene Delenia, Andrew Kellock and Kevin P. Roche for their help.

**Competing interests:** The authors declare no competing interests.

**Methods:**

**Magnetic measurements**:

The magnetic measurements of the Mn$_3$Ge film were conducted using a SQUID-VSM magnetometer by Quantum Design.

**Device sizes**:

The device size of the MTJs is estimated using $R_P$ and the value of resistance-area (RA) product determined from CIPT (Current In-Plane Tunneling) measurements. We find that these estimates agree well with the sizes obtained from TEM cross-sectional images.

**Film growth:**

The samples used in these studies were prepared using an ultra-high vacuum chamber with a base pressure of ~ $10^{-9}$ Torr. While the Ta and Mn$_3$Ge layers were deposited using ion beam deposition (with Kr gas) at a pressure $10^{-4}$ Torr, all the other layers were deposited by DC magnetron sputtering at an Ar sputter gas pressure of 3 mTorr (the MgO tunnel barrier was deposited by RF sputtering at the same pressure). All the films were deposited at room temperature.

**Write-error rate (WER) measurements**:

WER measurements were performed by connecting a Keithley 2602A multimeter and a pulse generator through a biased tee to the MTJ device. The 2602a unit was used for sending the reset pulse to set the initial state of the MTJ and reading the resistance of the MTJ. Either the Picosecond Pulse Generator 10070A or Tektronix AWG610 was used for sending the write pulse. The 10070A has a rise time of 55 ps and was used for the shorter writing pulses. Both the reset and write resistance states were measured after the state of the MTJ was set. $J_C$ for WER = 0.5 is obtained after accumulating the WER for different $J_C$ and using that data to obtain the fit for $J_C$ at WER = 0.5 (see Fig. 4b). For deeper error rate measurements (from Fig. 4c), measurement at a particular $J_C$ is run until any of these conditions are satisfied; either 10 errors are accumulated or $10^7$ trials have been performed. The WER then obtained is passed through the inverse of the standard normal cumulative distributive function and then linearly fitted against $J_C$ (solid lines) in Fig. 4c. The smallest $\varepsilon$ we can measure is only limited by measurement time.

**Figure 1.**

a

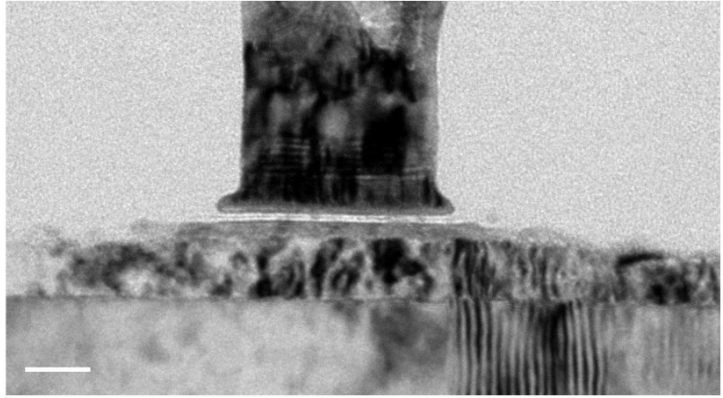

b
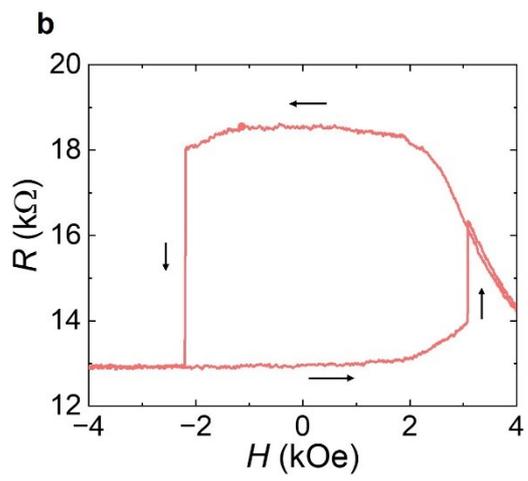

c
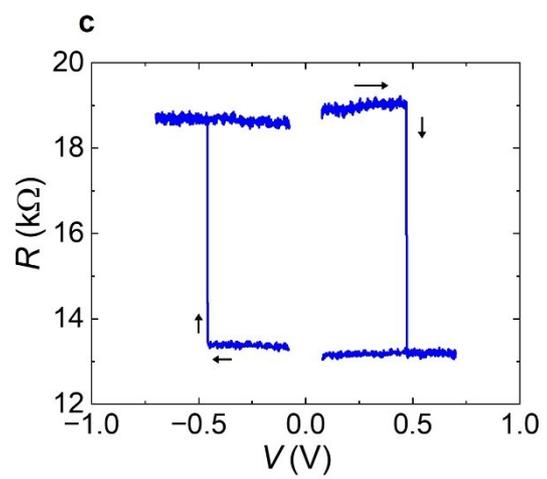

**Figure 2.**

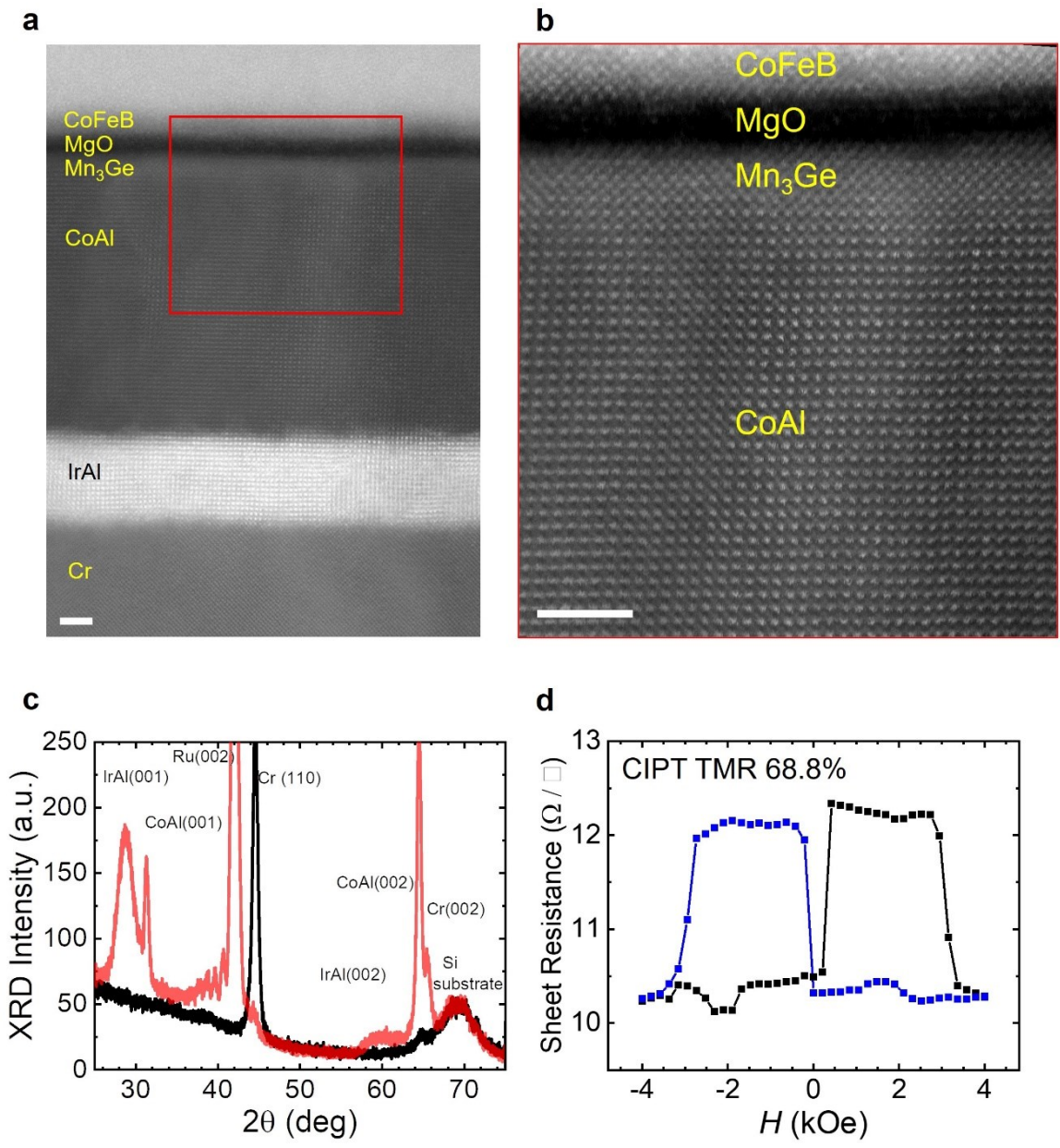

**Figure 3.**

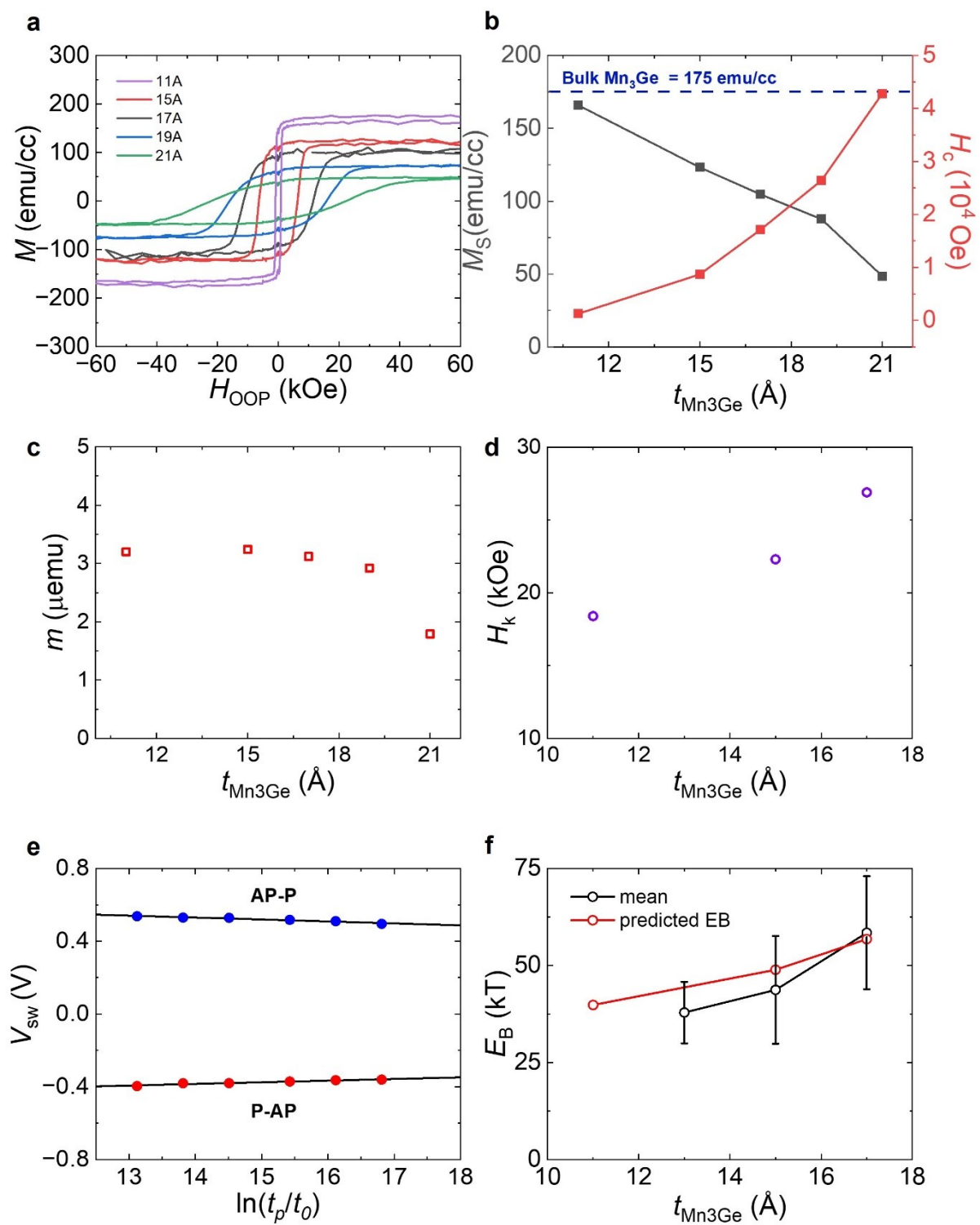

**Figure 4.**

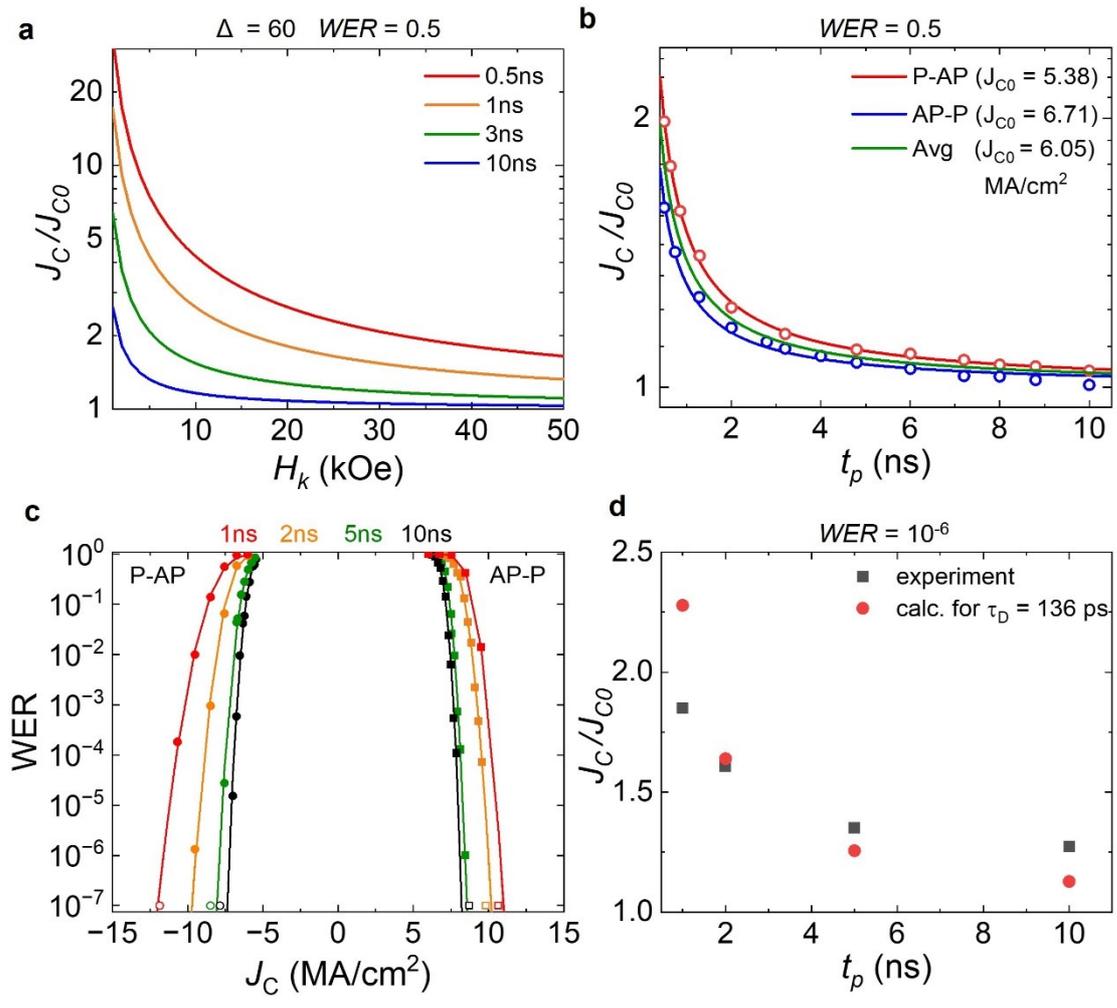

# Supplementary Information: Ferrimagnetic Heusler tunnel junctions with fast spin-transfer torque switching enabled by low magnetization

Chirag Garg[1+*], Panagiotis Ch. Filippou[1*], Ikhtiar[3*], Yari Ferrante[1], See-Hun Yang, Brian Hughes[1], Charles T. Rettner[1], Timothy Phung[1], Sergey Faleev[1], Teya Topuria[1], Mahesh G. Samant[1+], Jaewoo Jeong[3+,], and Stuart S. P. Parkin[2+]

[1]IBM Research - Almaden, San Jose, California 95120, USA.

[2]Max Plank Institute for Microstructure Physics, Weinberg 2, 06120 Halle (Saale), Germany.

[3]Samsung Semiconductor, Inc., San Jose, California 95134, USA

* These authors contributed equally to this work.
+ Emails of corresponding authors: chirag.garg1@ibm.com, j.jeong1@samsung.com, stuart.parkin@mpi-halle.mpg.de, mgsamant@us.ibm.com

**Supplementary Table 1: Exploration of different nitrides as the seed layer for the growth of $Mn_3Ge$ free layer magnetic tunnel junction**

| Nitride | TMR (%) | RA ($\Omega \cdot \mu m^2$) |
|---|---|---|
| ScN | 56.6 | 5.58 |
| TiN | 60.8 | 7.75 |
| VN | 58.8 | 7.13 |
| CrN | 51.9 | 4.88 |
| MnN | 62.3 | 19.4 |
| TaN | 56.9 | 9.11 |

Table shows current in-plane tunneling derived tunneling magnetoresistance (CIPT-TMR) values for $Mn_3Ge$ layer stacks grown on Si substrates utilizing the nitride/CTL (chemical templating layer) concept. These nitrides may not be stoichiometric. All film stacks are comprised of: 50 Ta/ 5 CoFeB/ Nitride / 400 Cr/ 50 IrAl/ 150 CoAl/ 13-19 $Mn_3Ge$/ 14-17 MgO/ 13-14.5 CoFeB/ 50 Ta/ 100 Ru, all thicknesses are in Å. $Mn_3Ge$ is annealed after deposition at ~390°C and there is a second annealing step at ~300°C after all layers are deposited to set the perpendicular magnetic anisotropy (PMA) in the CoFeB layer. The nitride thicknesses range from 1-10 Å, except for MnN which is 300Å thick.

**Supplementary Note 1: Selection and growth of suitable nitrides**

We show that nitrides extend the CTL technique of ordered growth of Heusler films from single crystal substrates to amorphous $SiO_x$. Here we discuss in detail the case of a metallic $Mn_xN$ layer (other nitrides are also similar). $Mn_xN$ films with different chemical compositions, varied from MnN to $Mn_{4.8}N$, were prepared by reactive magnetron sputtering using varying mixtures of Ar – $N_2$ sputter gas. This allows the lattice constant of the $Mn_xN$ layer to closely match that of the CTL.

We illustrate the growth of ultra-thin, 10 Å thick Heusler layers formed from $Mn_3Sb$, using a combination of $Mn_xN$ and CoAl CTL underlayers in Supplementary Figure 1. The detailed structure is as follows: Si(001)/ 250 $SiO_2$/ 50 Ta/ 3 CoFeB/ 300 $Mn_xN$/ 300 CoAl/ $Mn_3Sb$ or $Mn_3Ge$ / 20 MgO/ 20 Ta, all thicknesses are in Å. The variation of the lattice constant of the $Mn_xN$ layer as a function of nitrogen content is shown in Supplementary Figure 1a and b. $\theta$-$2\theta$ x-ray diffraction (XRD) scans in Supplementary Figure 1a show that the (002) $Mn_xN$ peak shifts to lower $2\theta$ angles with increasing nitrogen content. Thus, the $Mn_xN$ out-of-plane lattice parameter can be varied considerably from ~3.76 to ~4.26 Å with increasing nitrogen content, as summarized in Supplementary Figure 1b. When the nitrogen content is ~ 2.5 the lattice constant of the $Mn_xN$ layer matches closely with that of the CoAl CTL. However, we find that well-ordered CoAl can be prepared for a wide range of nitrogen content within the $Mn_xN$ layer. The chemically ordering within the CoAl layer gives rise to the (001) peak shown in Supplementary Figure 1a. As can be seen in the figure, a strong (001) peak, that varies little in $2\theta$ is observed for a wide range of x between ~2 and ~4. For the same range of nitrogen content, the surface of the film stacks was found to be very smooth. As can be seen in Supplementary Figure 1c, the root-mean-square roughness ($R_{rms}$) of the surface topography, found from atomic force microscopy (AFM) studies, was less than 3Å for x between ~2 and ~4.

Using this combination of $Mn_xN$ and CTL underlayers very thin Heusler films with excellent magnetic properties were obtained. Exemplary Magneto-optical Kerr effect (MOKE) perpendicular magnetic hysteresis loops are shown in the inset to Supplementary Figure 1d for 10 Å-thick $Mn_3Sb$ and 8 Å-thick $Mn_3Ge$ films. Both the $Mn_3Sb$ and $Mn_3Ge$ layers were deposited at room temperature (RT) but the $Mn_3Ge$ layer was in-situ annealed at ~ 340 ºC in ultra-high vacuum for 30 min before the capping layers were deposited. These both show excellent PMA with square hysteresis loops. This shows that the Heusler layers are highly thermally stable. The dependence of film coercivity ($H_c$) on the thickness of the Mn$_3$Ge layer ($t_{Mn_3Ge}$) is shown in Supplementary Figure 1d.

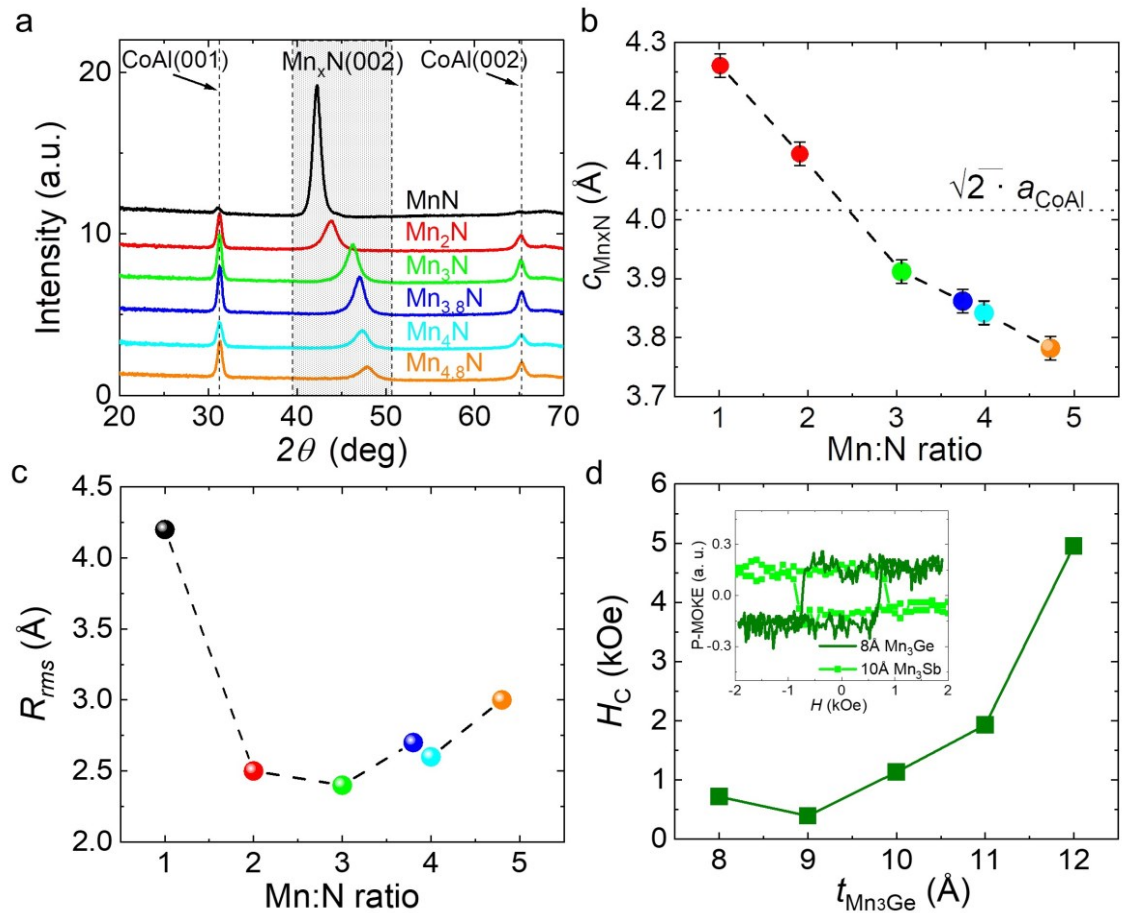

**Supplementary Figure 1.**

Growth of crystalline chemical templating layer on amorphous substrates. **a,** Out-of-plane θ-2θ XRD scans of CoAl films grown on Mn$_x$N with varying compositions. The dotted lines indicate the positions of the (001) and (002) CsCl-CoAl peaks. **b,** Dependence of the Mn$_x$N out-of-plane lattice constant on the Mn:N ratio extracted from (a). The dotted line indicates the value of the CoAl lattice constant after a 45° in-plane rotation. **c,** Surface roughness of the films shown in (a) as measured by atomic force microscopy. **d,** $H_c$ dependence of Mn$_3$Ge films with thickness $t_{Mn3Ge}$ grown on top of the optimized CoAl layer. Inset shows strong PMA in the magnetic hysteresis loops of 8 Å Mn$_3$Ge and 10 Å Mn$_3$Sb films measured using p-MOKE. The stack structure used for a-d is: Si(001)/ 250 SiO$_2$/ 50 Ta/ 3 CoFeB/ 200 Mn$_3$N/ 300 CoAl/ [Mn$_3$Ge or Mn$_3$Sb]/ 20 MgO/ 20 Ta, all thicknesses are in Å.

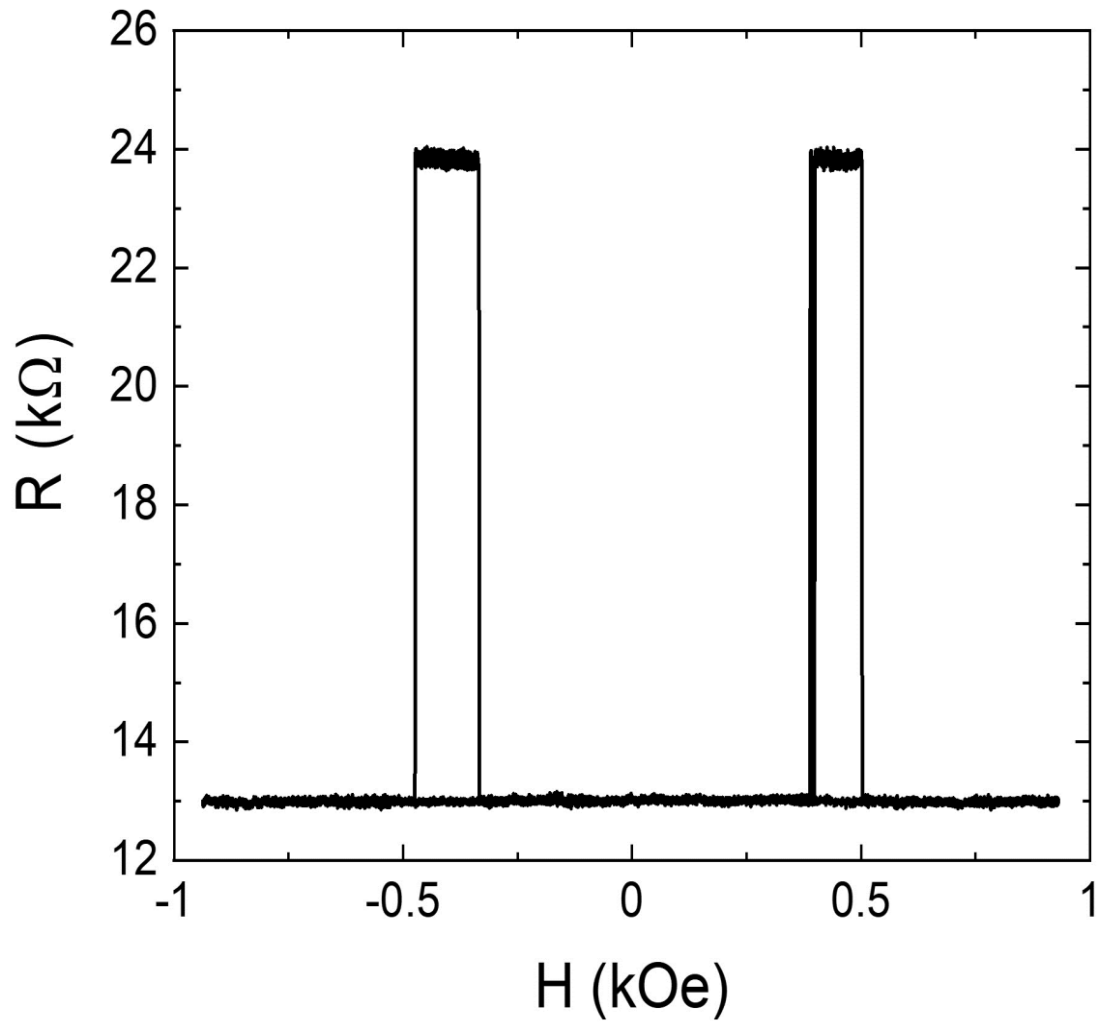

**Supplementary Figure 2:**

Resistance versus field (R-H), magnetic hysteresis loop for a 35 nm MTJ device with $Mn_3Ge$ free layer, showing a TMR of 87%. The stack is as follows: 50Ta/ 5CoFeB$_{20}$/ 300Mn$_x$N/ 400Cr/ 50IrAl/ 150CoAl/ 13Mn$_3$Ge/ anneal at ~340°C/ 17MgO/ 13CoFeB/ 50Ta/ 100Ru/ anneal at ~300°C, all thicknesses are in Å.

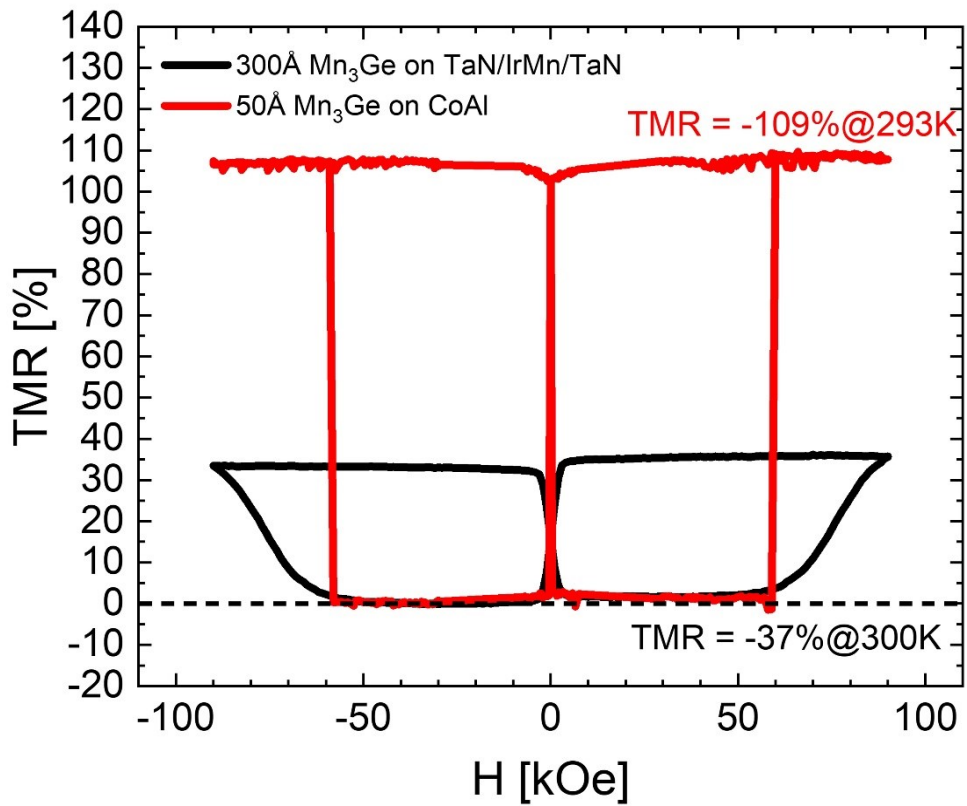

**Supplementary Figure 3:**

In red, high TMR from Mn₃Ge of 50Å where Mn₃Ge assumes its bulk properties of MnMn moment aligning with the total moment and TMR is negative. For comparison, in black, previously measured TMR on 300 Å bulk Mn₃Ge [1].

**Supplementary Note 2: Density functional theory calculations of Mn₃Ge**

In order to study the electronic structure and transport properties of Mn₃Ge deposited on CoAl substrate we performed density functional theory (DFT) calculations of the Mn₃Ge crystal structure with fixed in-plane lattice constant a = 4.03 Å (which equals to the lattice constant of CoAl) using the VASP program [2] with projector augmented wave (PAW) potentials [3,4] and Perdew-Burke-Ernzerhof (PBE) GGA/DFT functional [5]. We found that for fixed in-plane lattice constant a = 4.03 Å the relaxed out-of-plane lattice constant equal c = 5.964 Å (that corresponds to the dimensionless out-of-plane lattice constant c' = c/(2a) = 0.74). Note that a = 4.03 Å is just 1% smaller than the in-plane lattice constant $a_c$ = 4.06 Å of the cubic phase of Mn₃Ge [6] (that corresponds to the dimensionless out-of-plane lattice constant c' = c/(2a) = $1/\sqrt{2} \approx 0.707$). The convergence of the results was verified by varying the number of divisions in reciprocal space from 10×10×10 to 18×18×18 and the energy cutoff from 400 to 520 eV.

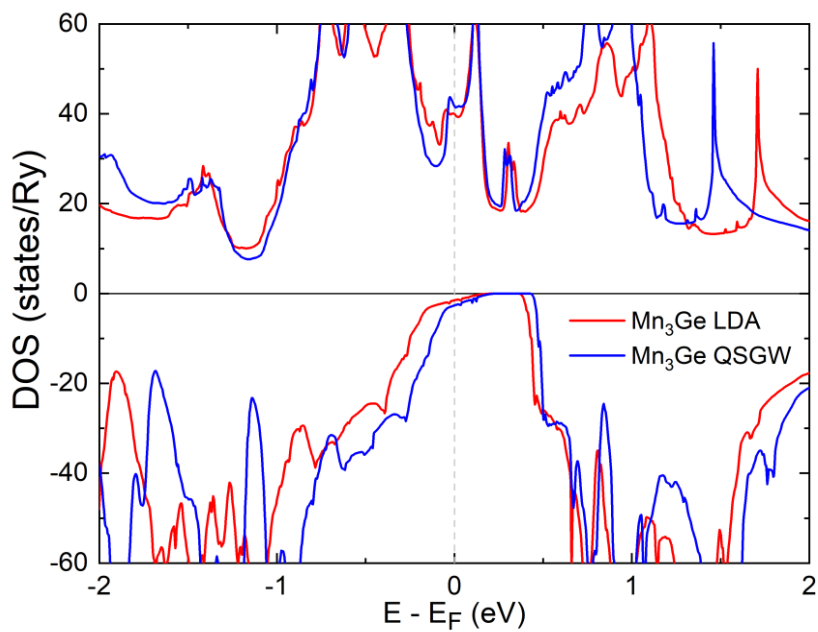

**Supplementary Figure 4:** The density of states of Mn₃Ge with in-plane lattice constant a = 4.03 Å calculated by LDA and QSGW methods.

The electronic structure of Mn₃Ge with in-plane lattice constant a = 4.03 Å and out-of-plane lattice constants c' = 0.74 was calculated using the full-potential all-electron linear muffin-tin orbital (LMTO) approach [7] with Barth-Hedin LDA/DFT functional [8], and also using the quasiparticle self-consistent GW (QSGW) method that is known to describe band gaps and other properties of materials with moderate *e-e* correlations significantly better than DFT [9–11]. The density of states (DOS) calculated by LDA and QSGW methods is presented Supplementary Figure 4. One can see that in both approaches the minority DOS has a valley near the Fermi energy resulting in large spin polarization (SP) of $Mn_3Ge$ for a = 4.03 Å. In particular, the spin polarization obtained by LDA is $SP_{LDA}$ = 0.92, and spin polarization obtained by more accurate QSGW method is $SP_{QSGW}$ = 0.88. The magnetic moment was found to be 1.02 $\mu_B$ in LDA and 1.03 $\mu_B$ in QSGW, that is close to the magnetic moment of $Mn_3Ge$ in cubic phase $m_c$ = 1.00 $\mu_B$ [6].

The transport properties of $Mn_3Ge$/MgO/Fe magnetic transport junction (MTJ) device with fixed in-plane lattice constant a = 4.03 Å were calculated using a tight-binding linear muffin-tin orbital method in the atomic sphere approximation (TB-LMTO-ASA) with the local density approximation of DFT for the exchange-correlation energy [12,13]. Relaxed positions of atoms at the $Mn_3Ge$/MgO interfaces (for both, the Mn-Mn and Mn-Ge terminations of the interface) were determined using the VASP molecular dynamic program [2]. The O-top configuration was found to be the most stable configuration (as compared with Mg-top and hollow) for both terminations at the $Mn_3Ge$/MgO interface (in agreement with Ref. [14]). For Fe/MgO interface the atomic positions from Ref. [15] were used.

Even though the $Mn_3Ge$/MgO interface can be very smooth (see, e.g., Ref. [1]) inevitably there will be atomic scale fluctuations in the morphology of the $Mn_3Ge$ layer that gives rise to regions with Mn–Mn and Mn–Ge terminations, due to the fundamental underlying structure of the Heusler compound. The simplest way to model such fluctuations is to average the transmission functions over the different terminations separately for parallel (P) and antiparallel (AP) configurations of magnetization of the $Mn_3Ge$ and Fe electrodes in the $Mn_3Ge$/MgO/Fe MTJ, assuming that the MgO thickness is the same across the device. The tunneling magneto resistance (TMR) is this simple model is calculated as TMR = $(T_P - T_{AP})/T_{AP}$, where transmission in the parallel configuration is given by $T_P = [T_{\uparrow\uparrow}(MnMn) + T_{\downarrow\downarrow}(MnMn) + T_{\uparrow\uparrow}(MnGe) + T_{\downarrow\downarrow}(MnGe)]/2$ and transmission in the antiparallel configuration is given by $T_{AP} = [T_{\uparrow\downarrow}(MnMn) + T_{\downarrow\uparrow}(MnMn) + T_{\uparrow\downarrow}(MnGe) + T_{\downarrow\uparrow}(MnGe)]/2$. (Here two arrows denote direction of the magnetization of $Mn_3Ge$ and Fe, correspondingly, and MnMn or MnGe denote the termination at the $Mn_3Ge$/MgO interface). The calculated TMR is shown in Supplementary Figure 5.

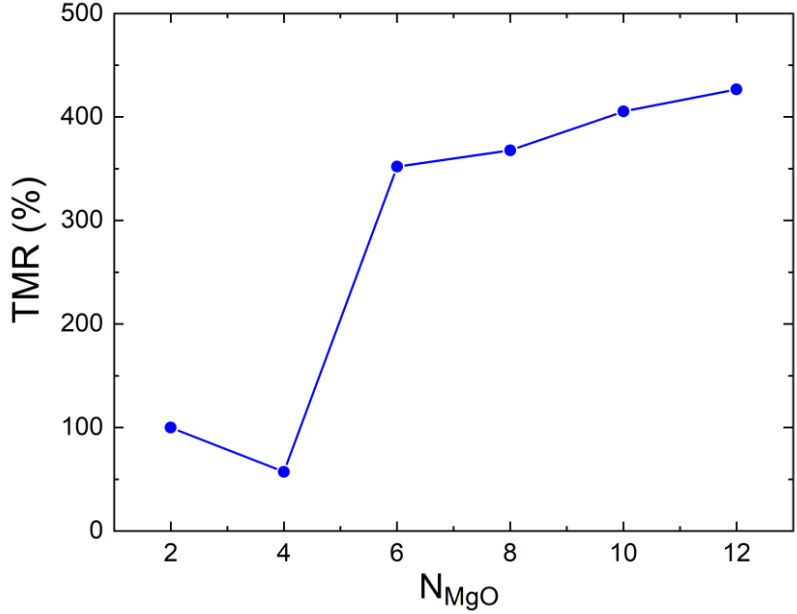

**Supplementary Figure 5**: TMR calculated for Mn$_3$Ge/MgO/Fe MTJ with in-plane lattice constant a = 4.03 Å and with an assumption of equal areas occupied by Mn–Ge and Mn–Mn terminations at the Mn$_3$Ge/MgO interface shown as a function of the number of MgO layers, $N_{MgO}$.

One can see that TMR is 100% for $N_{MgO}$ = 2 and 60% for $N_{MgO}$ = 4 and varies from 360% to 430% for $N_{MgO}$ ranging from 6 to 12. The high TMR values (~ 400%) at $N_{MgO} >= 6$ is a consequence of the high spin polarization of the near-cubic crystal structure of Mn$_3$Ge at a = 4.03 Å. Lower TMR values (~100%) at $N_{MgO} <= 4$ is a consequence of the presence of the interface resonance states localized near the Fe/MgO interface at the Fermi energy in Fe minority channel that leads to enhanced AP transmission (and therefore lower TMR) at small values of $N_{MgO}$.

**Supplementary Note 3: Extraction of anisotropy field magnetic properties of Mn₃Ge thin film**

The effective anisotropy field ($H_k$) of the Mn₃Ge film used in our study was estimated by obtaining the area enclosed (light green) between the OOP (black) and IP (blue) measurements of M-H (magnetization vs field). M-H was measured using a Quantum Design VSM-SQUID magnetometer capable of applying maximum field of 7 Tesla. The area enclosed gives the energy density difference between the OOP and IP configuration which when normalized by the saturation magnetization gives us the value of the effective anisotropy field. The Mn₃Ge films used for these measurements have been described earlier in the main text and their stack is: Si(001)/ 250Å SiO₂/ 1ScN/ 10CoAl/ '$t_{Mn_3Ge}$' Mn₃Ge/ 20 MgO/ 20 Ta, all thicknesses are in Å. For $t_{Mn_3Ge}$ = 11 Å, 15 Å, 17 Å, the M-H data and the enclosed curves illustrating our procedure are shown in Supplementary Figure 6a-c. We are not able to measure the anisotropy field for higher $t_{Mn_3Ge}$ as the field range of our magnet (7T) is not sufficient to completely saturate the magnetization during the IP measurement.

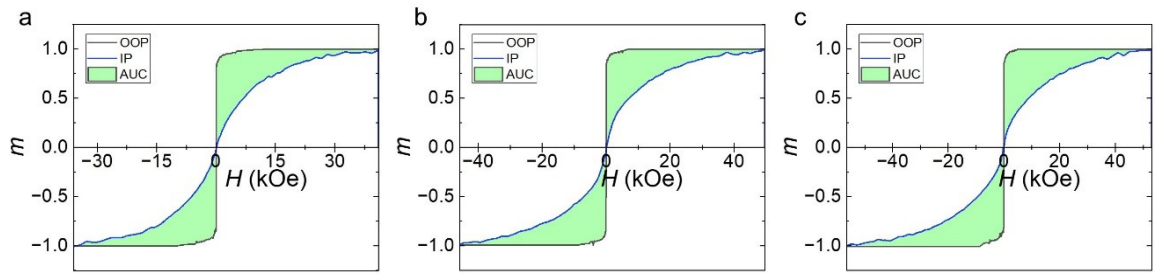

**Supplementary Figure 6**. Normalized magnetization ($m$) vs H curves for OOP (black) and IP configurations (blue). The enclosed area is shaded in light-green. (a-c) correspond to $t_{Mn_3Ge}$ = 15 Å b, $t_{Mn_3Ge}$ = 11 Å and c, $t_{Mn_3Ge}$ = 17 Å, respectively.